\documentclass[runningheads]{llncs}
\usepackage[margin=1in]{geometry}
\usepackage{hyperref}
\usepackage{orcidlink}

\renewcommand{\orcidID}[1]{\,\orcidlink{#1}}

\usepackage{wrapfig}
\usepackage{tabularx}

\usepackage[T1]{fontenc}
\usepackage[misc]{ifsym}
\usepackage{graphicx}
\usepackage{array}

\usepackage{amsmath}
\usepackage{amssymb}
\usepackage{multirow}
\usepackage{tikz}
\usetikzlibrary{positioning,backgrounds}
\usetikzlibrary{trees}
\usepackage{hyperref}
\usepackage{tablefootnote}

\usepackage{wasysym} 
\usepackage{ragged2e}
\usepackage{float}

\usepackage{adjustbox}
\newcommand{\headrow}[1]{\multicolumn{1}{c}{\adjustbox{angle=45,lap=\width-0.5em}{#1}}}
\newcommand{\flatrow}[1]{\multicolumn{1}{c}{{#1}}}

\newcounter{rq}
\renewcommand{\therq}{\arabic{rq}}

\newcommand{\RQ}[1]{%
  \refstepcounter{rq}%
  \par\smallskip
  \noindent\fbox{%
    \begin{minipage}{\dimexpr\linewidth-2\fboxsep-2\fboxrule\relax}
      \textbf{Insight \therq.} #1
    \end{minipage}}%
  \par\smallskip
}

\newcounter{gap}
\renewcommand{\thegap}{\arabic{gap}}

\newcommand{\gap}[1]{%
  \refstepcounter{gap}%
  \par\smallskip
  \noindent\fbox{%
    \begin{minipage}{\dimexpr\linewidth-2\fboxsep-2\fboxrule\relax}
      \textbf{Research Gap \thegap.} #1
    \end{minipage}}%
  \par\smallskip
}

\usepackage{xcolor}

\usepackage{xspace}

\newcommand{\etc}{\textit{etc.}\xspace}
\newcommand{\ie}{\textit{i.e.,}\xspace}
\newcommand{\eg}{\textit{e.g.,}\xspace}
\newcommand{\cf}{\textit{cf.}\xspace}

\newcommand{\circa}{\textit{ca.}\xspace}
\newcommand{\aka}{\textit{a.k.a.}\xspace}

\newcommand{\depm}{DePM\xspace}
\newcommand{\cepm}{CePM\xspace}
\newcommand{\depms}{DePMs\xspace}
\newcommand{\cepms}{CePMs\xspace}

\newcommand{\polyurl}[2]{Polymarket: `\href{#2}{#1}'}

\newcolumntype{L}[1]{>{\raggedright\arraybackslash}p{#1}} 

\begin{document}

\title{ SoK: Market Microstructure for Decentralized Prediction Markets (\depms)\thanks{This is a full version of the paper that appeared at the \textit{Workshop on Trusted Smart Contracts (WTSC}) at \textit{Financial Cryptography 2026}.}}

\author{
	Nahid Rahman\inst{1}\orcidID{0009-0001-9625-3816} \and 
	Joseph Al-Chami\inst{2}\orcidID{0009-0001-8381-8080} \and 
	Jeremy Clark\inst{1}\orcidID{0000-0002-3533-5965}
	}

\institute{
	Concordia University, Montreal, Canada \\ \email{nahid.rahman@mail.concordia.ca} \\ \email{j.clark@concordia.ca} \and
        Independent Researcher \\ \email{alchamijoseph@gmail.com}
	}
	
\maketitle


\begin{abstract}

Decentralized prediction markets (\depms) allow open participation in event-based wagering without fully relying on centralized intermediaries. We review the history of \depms which date back to 2011 and includes hundreds of proposals. Perhaps surprising, modern \depms like Polymarket deviate materially from earlier designs like Truthcoin and Augur v1. We use our review to present a modular workflow comprising eight stages: underlying infrastructure, market topic, share structure and pricing, initialization, trading, market resolution, settlement, and archiving. For each module, we enumerate the design variants, analyzing trade-offs around decentralization, expressiveness, and manipulation resistance. We also identify open problems for researchers interested in this ecosystem.

\end{abstract}


\section{Introduction}

In late 2024, the United States was in the midst of a presidential election when the decentralized prediction market, Polymarket~\cite{poly}, broke through mainstream news coverage~\cite{Cha24,YBG24a}. Stories focused, in particular, on the fact that it offered odds more favourable to eventual winner Donald Trump than those reflected in conventional polls and forecasts. Polymarket's odds are not set by experts or pundits, instead it is a specific type of betting market where odds are extrapolated from the prices of trades made in an open market (or somewhat open, as Polymarket was banned at the time in many countries including the US). 

As with traditional betting, whether online or through a bookie, prediction markets allow speculators to profit from correct forecasts~\cite{AFGH+08,WoZi06}. However the structure of a prediction market is different than traditional betting. One key difference is that prediction markets ease the process of moving in and out of bets before the event resolves, encouraging traders to place bets if they think the odds are over- or under-stated, and withdrawing profits if the odds realign.

It would be easy to think that Polymarket's design is the most obvious, straight-forward way to deploy a decentralized prediction market (\depm) on a blockchain. However the central thesis of this systemization of knowledge (SoK) paper is that Polymarket found success in bucking the trend. \depms were first given a few paragraphs in the Ethereum vision paper~\cite{But13}, released in late 2013 for the blockchain that would be deployed in 2015. Then two 2014 papers presented flushed out systems: a whitepaper called Truthcoin~\cite{Sz14} and an academic paper at WEIS 2014~\cite{BCFKMN14} (informally known as the `Princeton \depm' because of author affiliation). Developed independently,\footnote{The Princeton paper describes Truthcoin as being released while the paper was under review~\cite{BCFKMN14}, and the Truthcoin FAQ~\cite{Sz14b} mentions hearing about the Princeton paper but not having found the paper itself.} the two papers' designs are vastly different, representing two different goalposts for how a \depm might look. 

Early systems, like Augur~\cite{AKPWZ15} and Gnosis~\cite{Gn17} closely resembled Truthcoin~\cite{Sz14}, while modern systems like Polymarket~\cite{poly} either resemble the Princeton \depm~\cite{BCFKMN14} or use new solutions that resemble a hybrid of the two designs. Consider some examples:

\begin{enumerate}
\item From \S\ref{wf:price}, in Truthcoin, the market creator is active in setting initial prices (\ie odds) for each option and risks its own money (bounded)~\cite{Sz14}. In the Princeton system, the market creator is passive, not setting prices or risking any money~\cite{BCFKMN14}. Polymarket uses the latter~\cite{poly}. 

\item From \S\ref{wf:trade}, in Truthcoin, outcome shares are created with predecessor to an automated market maker (AMM)~\cite{Sz14}. In the Princeton \depm, outcome shares are traded with an orderbook~\cite{BCFKMN14}. Polymarket historically used an AMM and now uses an orderbook~\cite{poly}. 

\item From \S\ref{wf:close}, in Truthcoin, the blockchain decides event outcomes (\eg who won the election) through a reputation-based on-chain vote with slashing~\cite{Sz14}. In the Princeton paper, they are resolved through trusted arbiters acting as oracles~\cite{BCFKMN14}. The Ethereum whitepaper suggests both~\cite{But13}. In Polymarket, the third party oracle, UMA, operates under the hood through on-chain voting with slashing, but only when outcomes are disputed~\cite{poly,uma}.
\end{enumerate}

\paragraph{Research Questions.} Noticing these points of differences inspired us to enumerate all of the design decisions involved in creating a \depm. What are the core modules that appear across DePM designs? How do DePMs differ in how they implement each module (creation, trading, resolution, settlement)? Which early design choices shaped later DePM architectures? Which design choices create recurring failure modes or operational problems? What trade-offs do DePMs make between decentralization, usability, and correctness of resolution? We also ask what open design gaps remain for building robust DePMs (these are marked as `research gaps' in the paper).

\paragraph{Market Microstructure.} In traditional financial markets, the mechanics of trading are called market microstructure. Many market microstructure papers are empirical, studying how existing trading mechanisms affect prices, liquidity, spreads, trade execution, and other measures of market quality, rather than proposing new trading mechanisms. This is natural as traditional markets are mature and rarely redesigned. With \depms, by contrast, developers had an opportunity to build many of the basic mechanisms from scratch. Our SoK covers design options for different stages of \depms but the most depth is on microstructure topics, including how shares are structured, traded, and settled.

\paragraph{Systemization of Knowledge (SoK).} Within the blockchain literature, a variety of SoKs have been published with a variety of methodologies and deliverables, making it difficult to establish what constitutes an SoK paper. We review several SoKs from \textit{FC} and \textit{AFT} that, like us, consider a blockchain technology: oracles~\cite{ClEsGS21}, stablecoins~\cite{MSS20}, DeFi services~\cite{WPG+22}, light clients~\cite{CBC22}, layer 2s~\cite{GMR+20}, privacy technologies~\cite{BCDF23}, and MEV mitigations~\cite{HW22}. Like us, every one of these papers compares/contrasts a set of technologies~\cite{ClEsGS21,MSS20,WPG+22,GMR+20} or theoretical models~\cite{CBC22,BCDF23,HW22} from academia and industry. Like us, the main deliverable is decomposing designs into a taxonomy or classification system for comparison and evaluation~\cite{ClEsGS21,MSS20,WPG+22,GMR+20,CBC22,BCDF23,HW22}. Like us, some distill the technology into a single model into layers or abstractions before classifying~\cite{ClEsGS21}, others offer definitions~\cite{CBC22,WPG+22}. Only one of these papers does quantitative experiments~\cite{WPG+22}, a welcome but unnecessary requirement for SoKs. Unlike a literature review---which primarily summarizes prior work paper-by-paper---an SoK imposes a common set of concepts, abstractions, and evaluation criteria so that disparate systems can be compared on a level playing field rather than merely described.

\paragraph{Methodology.} We obtained a collection of academic works on decentralized prediction markets, as well as various intersecting topics including (centralized) prediction markets, oracles, DeFi, and AMMs. We used our knowledge of the field, Google Scholar (search and cited by features), and citations within papers. Our library is available, sorted by topic, on Zotero.\footnote{Zotero: \url{https://www.zotero.org/groups/5750510/2024_prediction_markets/library}} We also identified projects without white-papers or academic papers, reviewing the 97 current projects listed on a recently released online dashboard~\cite{Sal25} and by studying 20+ historical systems that are no longer active. We also searched news sources for opinions and issues on leading decentralized prediction markets, such as Polymarket. We applied the following exclusion criteria to remove from consideration designs that are: fully permissioned \cepms, alternative wager structures beyond the first notable design we saw (\eg parimutuel betting~\cite{ThZi88,KNZ08,ULSM25}, back/lay exchanges~\cite{CHKL21} and cash for difference contracts~\cite{ACES17}), supporting/adjacent infrastructure, and insufficiently documented or scams. We reviewed over a hundred projects, distilled into the 35+ notable \depms listed in Appendix~\ref{app:list}.

Our taxonomy primarily reflects the design landscape examined for the original SoK in 2026. We include selected subsequent developments and research findings where they illustrate important design trade-offs, but do not attempt to maintain a continuously updated survey of every platform.


A modular workflow (\cf~\cite{ClEsGS21}) is an abstraction that breaks competing designs into the same common set of steps and parts, enabling side-by-side comparison. We used multiple rounds of affinity diagramming to cluster \depm sub-components from our corpus into eight stages: underlying infrastructure, market topic, share structure and pricing, trading, market resolution, settlement, and archiving. Relevant artifacts and data are available on GitHub.\footnote{GitHub: \url{https://github.com/MadibaGroup/2024-PredictionMarkets}}

\section{Preliminaries}

\subsection{Wagering Systems}

Consider the following taxonomy of a few wagering systems:

\adjustbox{max width=\textwidth}{
\begin{tikzpicture}[
  font=\footnotesize,
  grow=right,                    
  level distance=40mm,           
  sibling distance=8mm,          
  box/.style={draw, rounded corners, align=left, inner sep=2pt, text width=28mm}
]
\node{Event wagering}
  child { node{Bookmakers} }
  child { node[]{Prediction markets}
    child { node[]{Centralized (\cepms) } }
    child { node[]{Decentralized (\depms)} }
  };
\end{tikzpicture}
}

\paragraph{Bookmakers versus prediction markets.}

A wager is a two-party contract with payouts based on the outcome of a future event. Consider Alice and Bob who wager on the same outcome of an event. With a fixed-odds bookmaker (or online betting), Alice's contract is different from Bob's contract in at least two regards: (i) it specifically names Alice as the counterparty and (ii) the payouts could be different if the odds changed between Alice's wager and Bob's. By contrast, in a prediction market contract (called a outcome share), Alice and Bob hold identical contracts: (i) all contracts are between the market operator and whoever redeems the contract, and (ii) the payout is exactly the same (typically \$0 if incorrect and \$1 if correct). Odds are reflected in the price paid for a prediction market contract (\ie variable cost and fixed payout), while a bookmaker contract has a fixed cost and variable payout. Thus the key distinction is that prediction market outcome shares are \textit{fungible} and can be freely traded between participants, enabling a free market that communicates information to the public through outcome share prices, trading volume, market depth, and other financial market metrics. Operating a prediction market also differs from operating a bookmaker in a very important way: 

\RQ{Bookmakers quote odds without needing to hold sufficient funds in the event of a payout. It is merely a promise to pay. By contrast, a prediction market is structured with sufficient collateral for all outcomes at all times (including the very first trade! See \S\ref{wf:init}).  Every dollar paid to a winning trader is a dollar from either (a) a losing trader or (b) bounded within a special loss fund setup before the market opens.}

\paragraph{\cepm versus \depm.}

The term \textit{decentralized prediction market} originates from the Ethereum whitepaper~\cite{But13} and we abbreviate it \depm to match terms like DeFi (decentralized finance)~\cite{WPG+22} and DePIN (decentralized physical infrastructure networks)~\cite{LWS+24,MEB+25}. The term \textit{decentralized}~\cite{OKK24} in each of these is actually shorthand for both \textit{decentralized} and \textit{permissionless}, where permissionlessness is generally the more important way \depms distinguish themselves from centralized prediction markets (\cepms). Permissionlessness could extend itself to the market topic, the trading of outcome shares, the closing of the market, and the withdrawing of rewards, but not all systems will open up each of these operations (as we will explain in \S\ref{sec:wf}). We say a system is \depm if at least one is permissionless.

 
 \section{Satoshi HBO Market}
\label{sec:hbo}

\begin{table}[t!]
\centering
\caption{Over a few days, truthful and untruthful (`cheap talk') evidence was presented to traders. The market reacted to correct signals and effectively filtered out fake signals, demonstrating a beneficial feature of prediction markets.\label{tab:hbo}}
\begin{tabularx}{\textwidth}{|L{1.2cm}|X|L{1.8cm}|L{1.8cm}|}
\hline
\textbf{Date} & \textbf{Information} & \textbf{Market Impact} & \textbf{Hindsight Verdict} \\ \hline
05 Oct & A long-dormant X account belonging to someone who had corresponded with Sassaman on Twitter posts a new message stating they were interviewed for the documentary. & Immaterial & Fake\\ \hline
05 Oct & Partially redacted leaked email from an HBO executive implies Len Sassaman. & Immaterial & Fake \\ \hline
07 Oct & CNN piece states director `confronts' Satoshi suspect `face-to-face' ruling out Sassaman, David Klieman, and Hal Finney. & Material & Truthful\\ \hline
07 Oct & Samson Mow, featured in the trailer, speculates it will name Adam Back. & Material & Wrong but factual basis \\ \hline
07 Oct & End credits of documentary leak featuring a tribute to Klieman. & Immaterial & Fake \\ \hline
07 Oct & Mow states Nick Szabo refused to discuss with director implying he was not `confronted'. & Material & Truthful \\ \hline
07 Oct & Widow of Sassaman states she was not interviewed. & Moderate & Truthful\\ \hline
08 Oct & A scene with Todd leaked but inconclusive if it is film's thesis. & Material & Truthful \\ \hline
08 Oct & Peter Todd confirms being confronted for documentary but unsure if he will be named. & Material & Truthful \\ \hline
08 Oct & Polymarket commenter claims screen test names Nick Szabo. & Immaterial & Fake \\ \hline
08 Oct & Fortune movie review discloses Todd is named & Very Significant & Truthful \\ \hline
08 Oct & Documentary airs and names Todd & Very Significant & Conclusive \\ \hline
\end{tabularx}
\end{table}

Before diving deep on the mechanics of decentralized prediction markets, we illustrate how markets work with a lighthearted example. On 3 Oct 2024, a trailer was released with press coverage of a new HBO documentary on Bitcoin to air about a week later on 8 Oct 2024. In an interview, the director stated, the film would question Satoshi's anonymous identity and, `who we land on is unexpected and is going to result in a fair amount of controversy.~\cite{Bec24}' The next day, Polymarket setup a market for speculating on who the documentary would name, providing 15 names plus an `other/multiple' option.\footnote{\polyurl{Who will HBO doc identify as Satoshi?}{https://polymarket.com/event/who-will-hbo-doc-identify-as-satoshi}} A benefit of a decentralized prediction market is allowing niche topics for markets, unlikely to attract mainstream betting websites---in this case, attracting \$44M USD in trading volume. Having an `other' option is also critical after many markets have failed to fully articulate every eventuality and in this case, the winner, was not one of the original 15 names (see Section~\ref{wf:topic}).

In game theory, \textit{cheap talk} describes strategic misinformation or signalling aimed at shaping beliefs or prices, provided the cost of deception is outweighed by the potential payoff~\cite{CrSo82}. This is well illustrated by what followed in the HBO Satoshi market as new pieces of evidence emerged, some real and some fake, with some fakes relatively elaborate (professional appearing end-credits or hijacking a target's X.com account) as summarized in Table~\ref{tab:hbo}. Further details are provided in Appendix~\ref{app:hbo}.

Also of interest is how the prediction market did not obviously extract \textit{insider information} which is in violation of what theory would predict~\cite{Han07}. The director did state he did not participate in the market and advised his team working on the film not to either~\cite{Beg24}. Friction for novice users is also high: web3 apps have a learning curve and if insiders were based at HBO in the US, access would require circumvention of Polymarket's geofencing. Perhaps these reasons kept insiders out of the market for the 5 days it ran. By contrast, allegations of insider trading have be levelled in other markets, including the removal of Nicolas Maduro from power immediately before his seizure by the United States military~\cite{Alm26}.

\gap{Can insider trades be classified from blockchain data with reasonable percision? Are occurrences higher on unregulated markets? Are markets with insider trading more accurate? What correlations exist between insider trading activity and market metrics (price, liquidity, \etc)?}

\label{app:hbo}

\begin{figure}
  \centering
  \includegraphics[width=\textwidth]{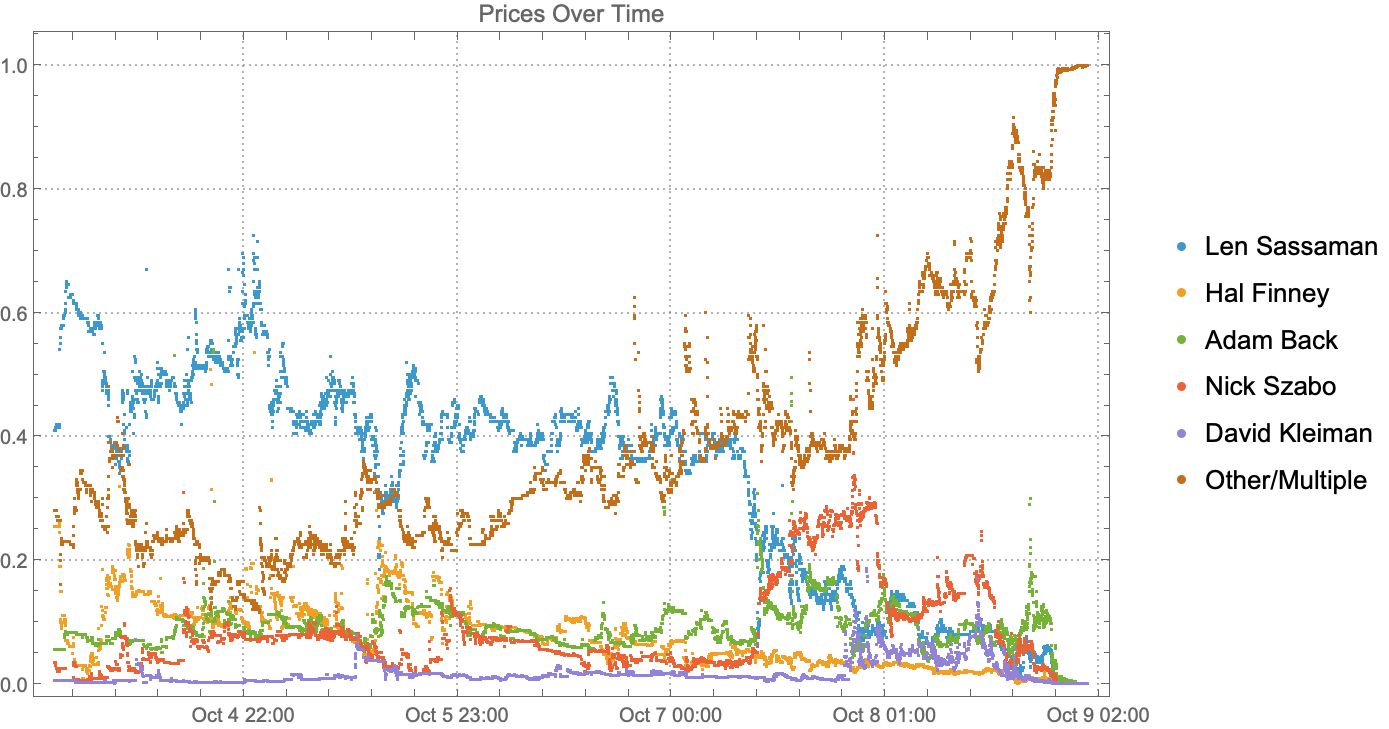}
  \caption{The price movements for 6 leading candidates in the Polymarket market for who would be named as Satoshi Nakamoto in the HBO documentary `Money Electric' which aired the evening of October 8.}
  \label{fig:example}
\end{figure}

\begin{figure}
  \centering
  \includegraphics[width=\textwidth]{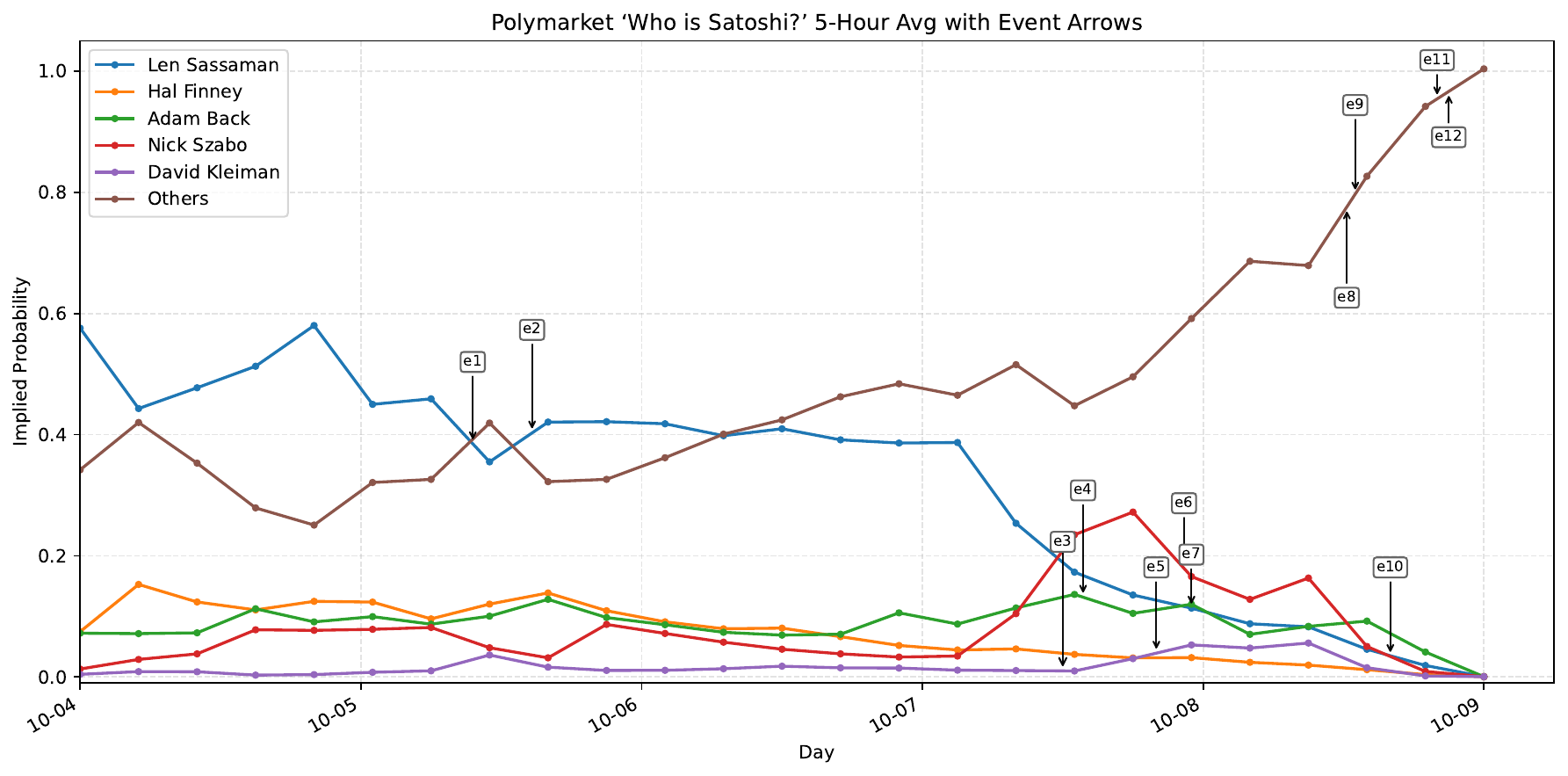}
  \caption{A variation of Figure~\ref{fig:example} with smoothing and super-imposed with the events (e1--e12) listed in Table~\ref{tab:hbo}.}
  \label{fig:example2}
\end{figure}


\section{Definitions}
\label{sec:defn}

We start with an abstract definition of a \depm market, premised on outcome shares ($j\in J$) being represented as fungible token types with standard notions of accounts, transfers, minting (issuance), and burning (redemption). We do not try to capture \cepms designs that are margin-based (\eg Kalshi~\cite{kalshi}) where traders hold only account-level positions, although such systems often satisfy analogous solvency and settlement conditions at the level of accounts rather than transferable tokens. Our definition generalizes on recent computer science-based definitions~\cite{BMR17,FPW23,SGKS25} which each presume specific sub-types of a prediction market (\eg \cite{BMR17,SGKS25} use \textit{merging/splitting} while \cite{FPW23} presumes \textit{automated bookmaking} from \S\ref{wf:trade}). 



We distinguish an \emph{event contract} from a \emph{trading venue}. An event contract specifies the outcome space and the payoff rules for a fixed set of fungible token types. A trading venue (e.g., an order book or AMM) is an implementation choice that quotes prices and executes trades in some subset of those token types. Throughout, we use `market' informally to refer to trading venues, and use `event contract' for the event-level specification defined below.

\begin{definition}[Event contract]\label{def:contract}
A (single) event contract is a tuple $C=(\Omega,J,R,t_{\mathrm{res}})$, where $\Omega$ is a nonempty outcome space, $J$ is a finite index set of fungible token types, $R:\Omega\to\mathbb{R}_{\ge 0}^{J}$ is a nonnegative payoff map, and $t_{\mathrm{res}}\in\mathbb{N}$ is the (scheduled) resolution time. Intuitively, $C$ is associated with an underlying uncertain event, whose possible resolutions are represented by $\Omega$.
\end{definition}

\paragraph{Remark: WTA special-case.}
For an event contract $C=(\Omega,J,R,t_{\mathrm{res}})$, suppose there exists a bijection $\iota:J\to\Omega$ and
$R(\omega)_j\in\{0,1\}$ with $\sum_{j\in J} R(\omega)_j=1$ for all $\omega\in\Omega$.
Then $C$ is a winner-take-all (WTA) or Arrow--Debreu contract: a unit claim of token $j$ pays $1$ iff the realized outcome equals $\iota(j)$, and $0$ otherwise.

\paragraph{Bundles and portfolios.}
A \emph{bundle} is a specified collection of token types within an event contract (for example, a set of tokens that the system treats as a unit for minting, burning, or conversion). A \emph{portfolio} is a vector $b=(b_j)_{j\in J}\in\mathbb{R}_{\ge 0}^{J}$ describing a trader's holdings, where $b_j$ is the number of units of token type $j$. The payoff of a portfolio at outcome $\omega\in\Omega$ is $\langle b, R(\omega)\rangle=\sum_{j\in J} b_j\,R(\omega)_j$. Bundles are used to define mechanisms such as splitting/merging and other payoff-preserving conversions; portfolios are the positions held by traders.

\begin{definition}[Prediction market system]\label{def:system}
A prediction market system is a tuple $\mathcal{S}=(\mathcal{C},\mathcal{N},\mathrm{res})$, where $\mathcal{C}$ is a countable set of event contracts, $\mathcal{N}$ is a numeraire (unit of accounting), and $\mathrm{res}$ is a resolution function over discrete time $t\in\mathbb{N}$ such that, for each $C=(\Omega,J,R,t_{\mathrm{res}})\in\mathcal{C}$, we have $\mathrm{res}(C,t)\in\{\bot\}\cup\Omega$ for all $t\in\mathbb{N}$, $\mathrm{res}(C,t)=\bot$ for $t<t_{\mathrm{res}}$, and there exists $\omega\in\Omega$ such that $\mathrm{res}(C,t)=\omega$ for all $t\ge t_{\mathrm{res}}$.
\end{definition}

\begin{definition}[System Axioms]\label{def:axioms} For every event contract $C=(\Omega,J,R,t_{\mathrm{res}})\in\mathcal{C}$ operating in system $\mathcal{S}=(\mathcal{C},\mathcal{N},\mathrm{res})$, the following axioms hold. In what follows, all quantities are for this fixed event contract $C$, and we write $\mathsf{Treas}:=\mathsf{Treas}_C$ and $\mathrm{L}:=\mathrm{L}_C$.

\begin{enumerate}

\item \textbf{Issuance and Solvency.}

Let $Q=(Q_j)_{j\in J}\in\mathbb{R}_{\ge 0}^{J}$ be the vector of outstanding supplies, where $Q_j$ is the total number of units of token $j$ that have been issued and not yet burned, and let $\mathsf{Treas}\ge 0$ be the event contract’s treasury (in the numeraire $\mathcal{N}$).

Define the worst-case liability
\begin{equation}
  \mathrm{L}(Q) \;:=\; \sup_{\omega\in\Omega}\; \sum_{j\in J} Q_j\,R(\omega)_j.
\end{equation}
The system maintains, at all times,
\begin{equation}
  \mathsf{Treas} \;\ge\; \mathrm{L}(Q).
\end{equation}
Any operation that changes supplies (including issuance) must preserve solvency. In particular, if issuance changes supplies from $Q$ to $Q'$, then the treasury is adjusted sufficiently:
\begin{equation}
  \Delta \mathsf{Treas} \;\ge\; \mathrm{L}(Q') - \mathrm{L}(Q).
\end{equation}

\item \textbf{Transfer and Fungibility.}
Holdings are transferable between accounts and transfers conserve per-token totals $Q_j$ (the system-wide outstanding supply of token $j$).

Units of each token type are fungible: for all $j\in J$, any $\omega\in\Omega$, and any $a\ge 0$, $a$ units of token $j$ entitle the holder to $a\,R(\omega)_j$ units of $\mathcal{N}$ if the realized outcome is $\omega$.

\item \textbf{Settlement and Redemption.}
If $t\ge t_{\mathrm{res}}$ and $\mathrm{res}(C,t)=\omega\in\Omega$, then any holder may redeem: for all $j\in J$ and $a\ge 0$, redeeming $a$ units of token $j$ pays $a\,R(\omega)_j$ units of $\mathcal{N}$. 
Redemption burns the redeemed tokens and debits the treasury by the corresponding payout, preserving the solvency invariant above.
 
 \end{enumerate}
 \end{definition}



\section{Modular Workflow}
\label{sec:wf}

%
%
%
%
%


\begin{table}[t!]

\caption{Classification of selected (for significance or novelty) \depms using our modular workflow. Keywords used in the table are defined in referenced section, while \CIRCLE,~\LEFTcircle, \Circle~respectively infer that the related property is fully, partially or not supported.\label{tab:class}}

\adjustbox{max width=0.9\textwidth}{
\begin{tabular}{|l|l|c|c|c|c|c|c|c|c|c|c|}

\flatrow{Year} & \flatrow{System}  & \headrow{Infrastucture (\S\ref{wf:chain})} & \headrow{Permissionless Topic (\S\ref{wf:topic})} & \headrow{Share Structure (\S\ref{wf:price})} & \headrow{Initial Issuance (\S\ref{wf:init})} & \headrow{Permissionless Trading  (\S\ref{wf:trade})} & \headrow{Resolution  (\S\ref{wf:close})} & \headrow{Settlement  (\S\ref{wf:clear})} & \headrow{Arch.  (\S\ref{wf:archive})} \\

\hline
2014 & Truthcoin~\cite{Sz14} & Sidechain & \CIRCLE & WTA & Bookmaking &\CIRCLE & Vote & Pull & \CIRCLE \\
\hline
2014 &Princeton~\cite{BCFKMN14} & Sidechain & \CIRCLE & WTA & Splitting & \CIRCLE & Arbiter & Push & \CIRCLE \\
\hline
2015 &Augur~\cite{AKPWZ15}  &  Smart Con. & \CIRCLE & WTA & Bookmaking &  \CIRCLE & Vote & Pull & \CIRCLE \\
\hline
2019 &Gnosis Sight~\cite{Bar19}  &  Smart Con. &\CIRCLE & WTA & Splitting & \CIRCLE  & Arbiter & Pull & \LEFTcircle \\
\hline
2019 &Augur v2~\cite{AKPWZ19}  &  Smart Con. &\CIRCLE & WTA & Splitting & \LEFTcircle, \CIRCLE  & Vote & Pull & \LEFTcircle \\
\hline
2020 &Polymarket~\cite{poly}  & Smart Con. & \Circle & YNB(-NR) & Splitting & \LEFTcircle, \CIRCLE & Opt., Vote & Gasless Pull & \LEFTcircle \\
\hline
2020 &Omen~\cite{Omen_docs}  &  Smart Con. & \LEFTcircle & YNB & Splitting & \CIRCLE & Opt., Vote & Push & \LEFTcircle\\
\hline
2021 &Kalshi~\cite{kalshi_docs} & Centralized  & \Circle & YNB & Matching & \Circle & Arbiter & Push & \Circle \\
\hline
2022 &Zeitgeist~\cite{zeit} & App-chain & \LEFTcircle & WTA & Bookmaking & \Circle & Opt., Vote & Pull & \LEFTcircle \\
\hline

\end{tabular}
}

\end{table} 

We now turn to the design landscape of \depms and step through our modular workflow (\cf~\cite{ClEsGS21}) of eight stages: underlying infrastructure, market topic, share structure and pricing, trading, market resolution, settlement, and archiving. For each stage, we enumerate the possible designs and discuss competing trade-offs, summarized as the columns of Table~\ref{tab:class}. 

\gap{Can accuracy metrics~\cite{GN07} be used to compare various \cepm and \depm platforms? Are there meaningful correlations between accuracy and the market microstructure options in Table~\ref{tab:class}?}


\subsection{Underlying Infrastructure}
\label{subsec:blockchain_infra}\label{wf:chain}

In theory, a decentralized and permissionless system might run on something other than a blockchain, but blockchain technology underlays all known \depms. The earliest (pre-Ethereum) research was in agreement that Bitcoin Script was not powerful enough to operate a \depm, and a \textit{sidechain}~\cite{BCD+14} would be required~\cite{Sz14,BCFKMN14}. Today, projects tend to run as \textit{smart contracts} on an Ethereum competitor (\eg Polygon~\cite{AKN21} or Solana~\cite{Yak18}) or an Ethereum L2 (\eg Arbitrum~\cite{KGCWF18} or Optimism~\cite{op}). There are no strong qualitative differences between the underlying blockchain---it is a choice driven by fees, user base, and supporting infrastructure. One other approach (\eg Zeitgeist~\cite{zeit} or SX Network~\cite{sx}) is to put the prediction market logic into the blockchain rules themselves on a custom \textit{app-chain}.

A near universal difference between any \cepm and \depm is that a \depm allows outcome shares to be withdrawn from the platform, typically in a form compliant with a token standard such as ERC-20 or the more efficient ERC-1155~\cite{LDBR24}. Withdrawing outcome shares allows traders to exchange tokens outside of the platform and to compose with third party DeFi services, including on-chain trading, lending, and leverage~\cite{WPG+22}.

Perhaps the biggest evolution in \depm design is infrastructure~\cite{Gun24}. Early \depms were monolithic, single-vendor codebases. Modern \depms are built from existing infrastructure, which composes through highly standardized interfaces. For example, Polymarket's original architecture was built on Gnosis' Conditional Tokens Framework (CTF)~\cite{ctf}. The numeraire is Circle's USDC stablecoin, which can be bought with a credit card through MoonPay. Trading outcome shares and USDC works out-of-the-box on any platform (on- or off-chain) that supports ERC-1155 tokens. Market outcome disputes are escalated to UMA's DVM oracle~\cite{uma}. Polymarket also uses third party services for bridging assets, embedded wallets (based on email verification), and EIP-3009 gasless withdrawals. Beyond software engineering benefits, building a service by composing modules can enhance trust agility, which is the ability to quickly swap out modules that are faulty or malicious. For example, Polymarket could switch from UMA to say Chainlink or Kleros, with less effort than if the oracle service was vertically integrated. 

Polymarket subsequently began upgrading its architecture in 2026, introducing new exchange infrastructure, a new collateral token, and its own position-token framework~\cite{+polyclob,+polyv2}. Unless otherwise noted, our descriptions of Polymarket's mechanisms refer to its original CTF-based architecture.

\gap{On-chain markets provide open information about the trading strategies of other traders who do not deliberately and successfully obfuscate their trades and identities. Can we classify copy traders and measure occurrences? Is copy trading profitable? Under what conditions? Is there a negative impact on market accuracy due to cascading/herding effects? Update: Subsequent empirical studies have begun addressing these questions by identifying copy trading on Polymarket, measuring its profitability, and examining the practical limitations of following successful traders~\cite{Oli26,JiSu26}.}




\subsection{Market Topic}\label{wf:topic}



\begin{table}[t!]

\caption{Some pitfalls that illustrate the difficulty in properly defining a prediction market topic, collected through our literature review and manual observation of top markets on Polymarket over 2024--2025. \label{tab:pitfalls}}

\adjustbox{max width=\textwidth}{
\begin{tabular}{L{0.175\textwidth} L{0.8\textwidth}}
\hline
\textbf{Pitfall} & \textbf{Description} \\
\hline


\multirow{2}{\linewidth}{Borderline Categories}
  & \textit{Example}: A market on whether Zelensky would wear a \textit{suit} was contested when he wore a single-breasted jacket with patch chest pockets and matching trousers;\tablefootnote{\polyurl{Will Zelenskyy wear a suit before July?}{https://polymarket.com/event/will-zelenskyy-wear-a-suit-before-july}} media equivocated on describing it as a suit.\tablefootnote{Google Docs: \href{https://docs.google.com/document/d/1p0CSpse6YwLApvwKt173bDg1cQVNcNEe0_2sNCvhaZs/}{Did President Zelenskyy wear a suit before July 2025?}.} Other ambiguities include whether enforcement against TikTok in the US constitutes a \textit{ban},\tablefootnote{\polyurl{TikTok banned in the US before May 2025?}{https://polymarket.com/event/tiktok-banned-in-the-us-before-may-2025}} or if finding debris from the Titan submersible constitutes it being \textit{found}.\tablefootnote{\polyurl{Will the missing submarine be found by June 23?}{https://polymarket.com/event/will-the-missing-submarine-be-found-by-june-23}} \\ \cline{2-2}
  & \textit{Mitigation}: Clearly state inclusion/exclusion criteria (\eg a subsequent market on a potential hug between Trump and Putin spent a paragraph defining a hug.\tablefootnote{\polyurl{Will Trump and Putin hug on Friday?}{https://polymarket.com/event/will-donald-trump-and-vladimir-putin-hug-on-friday}}) \\ \hline
  

\multirow{2}{\linewidth}{Precedence Gaps}
  & \textit{Example}: A proposition bet on the colour of the 2014 Super Bowl  `Gatorade shower' was contested when the coach was showered twice with different colours~\cite{BCFKMN14}. A market on whether Zelensky would be `the' 2022 TIME Person of the Year was contested when both Zelensky and the Spirit of Ukraine were named.\tablefootnote{\polyurl{Will Volodymyr Zelenskyy be the 2022 TIME Person of the Year?}{https://polymarket.com/event/will-volodymyr-zelenskyy-be-the-2022-time-person-of-the-year}} \\ \cline{2-2}
  & \textit{Mitigation}: Parse the predicate for any statements needing explicit precedence (\eg first, majority, primary); or establish a payout rule for ties; or include an outcome for `multiple.' \\ \hline


\multirow{2}{\linewidth}{Hidden Presumptions}
  & \textit{Example}: A market concerning a divorce presumes the couple are married (as opposed to common law) which was unknown.\tablefootnote{\polyurl{Astronomer Divorce Parlay}{https://polymarket.com/event/astronomer-divorce-parlay}} \\ \cline{2-2}
  & \textit{Mitigation}: Parse the predicate for any presumptive statements and remove/address them.\\ \hline


\multirow{2}{\linewidth}{No Ground Truth}
  & \textit{Example}: A market on whether a US strike destroyed an Iranian nuclear facility was contested when each country reported different outcomes and no neutral third party was granted access to the site.\tablefootnote{\polyurl{Fordow nuclear facility destroyed before July?}{https://polymarket.com/event/fordow-nuclear-facility-destroyed-before-july}} A market on whether Baron Trump was `involved' in the \$DJT memecoin lacked an authoritative source.\tablefootnote{\polyurl{Was Barron involved in \$DJT?}{https://polymarket.com/event/was-barron-involved-in-djt}} An election market on Venezuela's president was contested when the government declared Maduro won, while international media and democracy watchdogs declared Gonzalez received more votes.\tablefootnote{\polyurl{Venezuela Presidential Election Winner}{https://polymarket.com/event/venezuela-election-winner}} \\ \cline{2-2}
  & \textit{Mitigation}: Avoid markets without ground truth sources; or include an additional option in the market for unverified. \\ \hline


\multirow{2}{\linewidth}{Platform Coupling}
  & \textit{Example}: Hypothetically, traders who correctly predict USDC will completely de-peg on a platform that pays out in USDC will receive a payout but it will be worthless (\cf~\cite{BCFKMN14}). \\ \cline{2-2}
  & \textit{Mitigation}: Avoid markets that are self-referential, including topics on the platform itself and its numeraire. \\ \hline


\end{tabular}
}
\end{table}


\cepms include the Iowa Electronic Markets~\cite{iem}, Kalshi~\cite{kalshi_docs}, and PredictIt~\cite{predictit_faq}, as well as InTrade historically~\cite{intrade}. These systems exercise control over what topics may form a market and thus are \textit{permissioned} with respect to market topics. They also operate under regulations that may restrict markets to certain topics or fully ban operations in regulated jurisdictions~\cite{Du19,Ma24}. 

By contrast, \depms like Augur~\cite{AKPWZ15,AKPWZ19}, the original Gnosis~\cite{Gn17}, and PlotX~\cite{plotx_docs} enable \textit{permissionless} market creation by any user without centralized review. This removes the regulatory hook, enables niche topics that might not attract mainstream interest~\cite{WZ24}, and allows markets to be created without delay after real world events. However it can also lead to a greater incidence of malformed (or even malicious) market definitions, spam duplicates of existing markets, and unlawful topics, such as the `assassination markets' which appeared on Augur in 2018~\cite{Du19}. \depms are generally web3 applications which means that a web-based user interface mediates transactions between the user and the underlying smart contracts. Market topic moderation could be implemented at the web3 layer (\eg Predictions.Global unlisted assignations markets from Augur's smart contracts~\cite{Du19}) but this does not prevent users from building an alternative UI or directly transacting with the smart contracts. 


These systems are not all purely permissionless. A \textit{hybrid model} puts some controls on topic creation without centralizing it fully~\cite{AKPWZ19}. For example, proposers may have to stake tokens to propose a market, and while the market is optimistically published, a review (either centralized or via an on-chain voting mechanism) could remove the market and/or slash the proposer.  

If issues in a market's topic or definition are uncovered while the market is still active, \depms like Polymarket allow `additional context' notes to be added. However these clarifications could alter the market ex post and also disadvantage traders who do not see the note. The latter can be mitigated by advertising that a note will be published, always publishing at the same time (\eg 5pm ET), and clearing standing limit orders from an orderbook before posting~\cite{poly}.

\RQ{Careful attention must be paid to both the general topic of the market and the `fine-print' or exact predicate that decides the market. Table~\ref{tab:pitfalls} provides several examples of pitfalls. Dealing with definitional pitfalls has been, to date, a trial and error process where market creators learn from past mistakes and ad hoc `legalese' (e.g., a `consensus of credible reporting' may be used to resolve markets) is copied from market to market~\cite{Ad24}.}

\gap{Can market topics be written with machine-checkable predicate specifications (precedence rules, ranked sources, time semantics, and default outcomes), where model checking (\cf~\cite{Cla21}) could eliminate ambiguities, corner cases, and loopholes? What would real world case studies of markets with additional context tell us?}


\subsection{Share Structure and Pricing}\label{wf:mech}\label{wf:price}

The core requirement of a prediction market is that wagers are represented by fungible outcome shares. The structure of outcome shares typically falls into one of three categories and two variants (although more exotic structures are possible and explored in research). 

\subsubsection{WTA.}

The first structure we term \textit{winner-take-all (WTA)} (\aka \textit{categorical}) and was popularized by Iowa Electronic Markets~\cite{iem}. Consider a market with three possible outcomes: $\Omega=\{\mathsf{A}, \mathsf{B}, \mathsf{C}\}$.  A WTA market issues an outcome share for each outcome $J=\{j_\mathsf{A}, j_\mathsf{B}, j_\mathsf{C}\}$. If there are only two outcomes, it is called a \textit{binary market}. If the outcome $\omega$ is $\mathsf{B}$, the share $j_\mathsf{B}$ pays \$1 (or one unit of numeraire $\mathcal{N}$) and the other shares pay \$0.  For any $k\in\{\mathsf{A},\mathsf{B},\mathsf{C}\}$,

\begin{equation}
R(\omega)_{j_{k}}=\begin{cases}1, & \text{if } \omega=k,\\0, & \text{otherwise.}\end{cases}
\end{equation}

For a WTA market to be well-functioning, two conditions must hold on outcome shares. (i) They should be \textit{mutually exclusive} so no more than one share wins: $R(\omega)_{j_{k}}\,R(\omega)_{j_{\mathsf{\ell}}}=0 \quad \forall\omega\in\Omega, \forall k\neq\mathsf{\ell} $; and (ii) they should be \textit{complete} so at least one share wins: $\sum_{k\in\Omega} R(\omega)_{j_{k}}=1\quad \forall \omega\in\Omega$. If they are not mutually exclusive, the operator could be under-collateralized for making all payments (in violation of solvency in Definition~\ref{def:axioms}). If they are incomplete, a deficient market might end with all participants receiving \$0. A consequence is that holding one share for each outcome is equivalent to holding \$1, a fact we will return to in \S\ref{wf:trade}. 

In a WTA market, the price of a outcome share (\eg $p(j_\mathsf{A})=\$0.54$) is a proxy for the probability that the outcome will occur (\eg $\mathrm{Pr}[\omega=\mathsf{A}]=54\%$). A common adage is the prices of each share sum to \$1.00 ignoring fees and discounting (\eg $p(j_\mathsf{A})=\$0.54$, $p(j_\mathsf{B})=\$0.23$, $p(j_\mathsf{C})=\$0.23$) but this is imprecise~\cite{BCFKMN14}. Outcome shares (like anything) have two prices: a bid price (what a trader is willing to buy for) and an ask price (willing to sell for). If the sum of the bid prices exceeds \$1.00 or if the sum of ask prices are below \$1.00, arbitrageurs have an opportunity to secure risk-free profit through a trade that will erase the condition when fully extracted. This means the sum of bids and sum of asks should result in the bid-ask spread straddling \$1.00 but the amount of the spread could be arbitrarily large. So in user interfaces that display a single `price' (\eg the last sale price or the midpoint between the best bid and the best ask), prices may indeed not sum to \$1.00---this is not a market failure, just a misunderstanding. 

 
 \subsubsection{YNB.}
 
The second structure we term a \textit{yes-no bundle (YNB)}. YNB markets were popularized by InTrade~\cite{intrade}. A YNB market issues two token types for each outcome, a `yes' and a `no:'  $J=\{j_{\mathsf{A_Y}}, j_{\mathsf{A_N}}, j_{\mathsf{B_Y}}, j_{\mathsf{B_N}}, j_{\mathsf{C_Y}},j_{\mathsf{C_N}} \}$. For any $k\in\{\mathsf{A},\mathsf{B},\mathsf{C}\}$,

\begin{equation}
R(\omega)_{j_{\mathsf{k_Y}}}=
\begin{cases}
1, & \text{if } \omega=k,\\
0, & \text{otherwise.}
\end{cases}
\qquad
R(\omega)_{j_{\mathsf{k_N}}}=
\begin{cases}
1, & \text{if } \omega\neq k,\\
0, & \text{otherwise.}
\end{cases}
\end{equation}

Each outcome-specific pair $\{j_{\mathsf{k_Y}}, j_{\mathsf{k_N}}\}$ constitutes a binary WTA market ($k$ vs.\ not-$k$). A YNB market is the union of these pairs, so the WTA exclusivity and completeness properties hold per pair. However exclusivity and completeness do not necessarily hold across all bundles, allowing more flexible markets. For example, a market on what words Trump will say in a congressional address included Bitcoin (no), beautiful at least 10 times (yes), and Canada (yes).\footnote{\polyurl{What will Trump say during address to Congress?}{https://polymarket.com/event/what-will-trump-say-during-state-of-the-union?tid=1755462445353}} Multiple words can resolve to yes (not exclusive) and it is possible he says none of the listed words (not complete). An established YNB market that is being actively traded, because it is not complete, can have new outcomes added to it fairly, whereas WTA markets must account for every possible outcome at the start of the market (perhaps utilizing an `other' outcome).

\subsubsection{YNB-NR.}

A variant of the YNB market is one where, even though it is not necessary, the yes share outcomes are in fact complete and exclusive. In other words, each yes/no bundle is a WTA market and the set of all yes shares is also a WTA market. We term this YNB variant as \textit{negative risk} (YNB-NR), a term introduced by Polymarket~\cite{poly}. Recall that in a WTA market, roughly speaking, the share prices sum to \$1 (modulo the fine print about bid/ask spreads above). For a YNB-NR market, the Yes shares sum to \$1, while the No shares will sum to $|\Omega|-1$.

\begin{equation}
\sum_{k\in\Omega} R(\omega)_{j_{\mathsf{k_Y}}}=1
\quad\text{and}\quad
\sum_{k\in\Omega} R(\omega)_{j_{\mathsf{k_N}}}=|\Omega|-1
\quad\text{for all }\omega\in\Omega.
\end{equation}



%
%

Polymarket introduced a \textit{negRisk} gadget that allows a trader holding a portfolio of No shares across $A$ outcomes ($|A|>0$) to extract cash by converting to the complimentary Yes shares:

\begin{equation}
\sum_{k\in A} R(\omega)_{j_{k_N}}
=
(|A|-1)
+
\sum_{\ell\in\Omega\setminus A} R(\omega)_{j_{\ell_Y}} 
\quad\text{for all }\omega\in\Omega\text{ and any }A\subseteq\Omega.
\end{equation}

\paragraph{Example.} Alice owns 1000 No shares for Adam Back, Hal Finney, and Nick Szabo in a market about an HBO documentary set to identify Satoshi Nakamoto. Since only one share type can win, Alice is guaranteed at least \$2000 from her positions at resolution (and up to \$3000 is she is right about all three). It is capital inefficient to keep her \$2000 locked up in her portfolio of No shares. NegRisk allows her to converts her position into 1000 Yes shares for every other candidate and extract the \$2000 in cash, maintaining the exact same exposure to the market (\$1000 at risk).

\subsubsection{Scalar.}

The third structure is a market where the outcome is a quantity of interest (\eg popular vote, temperature, price level, \etc) observed at a cutoff time with a strict lower bound and upper bound. Termed a \textit{linear} or \textit{scalar} market, there is only one share and its payout is what value the quantity takes on (typically normalized to the range $[0,1]$ with rounding). As an example, in a market on Trump's popular vote, if the quantity is 49.8\%, the share will pay \$0.498. Shares can also be sold in bundles with `long' receiving \$0.498 and `short' receiving (\$1-\$0.498).

Formally, if we let $X:\Omega\to\mathbb{R}$ be the observed quantity, and $[a,b]$ be an interval of values, then the linear outcome share $j_{\mathrm{lin}}$ pays:

\begin{equation}
R(\omega)_{j_{\mathrm{lin}}}=
\begin{cases}
0, & X(\omega)\le a,\\
\dfrac{X(\omega)-a}{\,b-a\,}, & a< X(\omega) < b,\\
1, & X(\omega)\ge b.
\end{cases}
\end{equation}

While scalar markets are supported by \depms like Augur~\cite{AKPWZ15,AKPWZ19} or those based on Gnosis' Conditional Tokens Framework (CTF)~\cite{ctf}, including Polymarket~\cite{poly} and Omen~\cite{Omen_docs}, they are not frequently used. For Polymarket, markets instead cheat, approximating a scalar market  by splitting the quantity into `buckets' and running a YNB market for each bucket. This avoids a less-vetted codebase within CTF, unifies the user interface across market types, and possibly avoids small edge cases over the exact resolution of the quantity (\eg off by 0.1 percentage disputes). However a problem with buckets is as follows: Alice estimates correctly that Trump will win the election with 49--51\% of the popular vote. If there is a bucket for 45--49.9\% and a bucket for 50--54.9\%, Alice's forecast does not fit into a single bucket. Alice buys both buckets, knowing only one will win, diluting her expected return on capital. A second consequence of bucketization is volatile market jumps when the market consensus crosses from one expected bucket into a neighbouring bucket.\footnote{\polyurl{April 2025 Temperature Increase (ºC)}{https://polymarket.com/event/april-2025-temperature-increase-c4}}  

\gap{How does a \textit{negRisk} gadget encourage price parity between equivalent positions? Can we quantify its improvements for price execution and trade execution costs? Does a reverse gadget, for converting a single yes share, offer similar benefits? Update: subsequent research has investigated these questions empirically and implemented a prototype supporting reverse negRisk conversions~\cite{GMM26}. The results highlight a distinction between portfolios that are equivalent at settlement and those that can be converted into one another before settlement.}

\gap{The academic literature has explored advanced market types, including combinatorial markets and establishing conditional probabilities from markets. Implementing and experimenting with these designs for \depms is open. In sports-betting, chaining together multiple outcomes in an all-or-nothing bet (parlay) is popular but a treatment for \depms is lacking. }


\subsection{Market Initialization}\label{wf:init}

Options for trading outcome shares can be broken into two steps: (i) how does the first outcome share come into existence and how does the first trader trade, and (ii) how do traders trade once a market has been established? Probably the greatest evolution in \depms, from Truthcoin to Polymarket, concerns how the first trade happens. There are three options: \textit{automated bookmaking}, \textit{splitting}, and \textit{matching}. 

\textit{Automated bookmaking} was popularized through the academic work of Robin Hanson~\cite{Ha03,Han07b}, researched for Intrade~\cite{McC08}, and first suggested for \depms by Truthcoin~\cite{Sz14}, which heavily influenced Augur v1~\cite{AKPWZ15}. 

\begin{definition}[Automated bookmaking]\label{def:autobook}
Fix a market $M=(\Omega,J,R,t_{\mathrm{res}})$. An \emph{automated bookmaker} posts (i) a loss fund $\mathsf{Treas}\in\mathbb{R}_{\ge 0}$ in the numeraire $\mathcal{N}$ and (ii) a pricing rule (odds) for token types $j\in J$. It accepts deposits in $\mathcal{N}$ and issues shares, updating the supply vector $S$, and must always satisfy solvency (Definition~\ref{def:axioms}). 
\end{definition}

In this model, the operator sets initial prices for each outcome share (equivalent to setting market odds) and collateralizes enough payout money to cover a worst-case loss in its treasury. If Alice is the first trader, she can immediately trade with the operator. The operator is autonomous and sets buy/sell prices algorithmically, originally using Hanson's logarithmic market scoring rule (LMSR)~\cite{Han07b}. The key point is that the operator's fund is Alice's counterparty; if Alice wins, the fund loses, and vice-versa. The pros are instant liquidity for the first trader and the cons are the risk to the operator of losing money and the burden of needing to set initial odds (getting them wrong increases the chances it loses). 

Acute readers might wonder the relationship between automated bookmaking and  an automated market makers (\eg Uniswap), which are discussed in the next section. Roughly speaking, a WTA market run by automated bookmaking is termed a cost-function prediction market (CFPM) and a CFPM is equivalent in pricing (and trade costs) to an AMM (defined by a set of axioms) with the right invariant~\cite{FPW23}.

The second approach, \textit{splitting}, was first suggested for \depms by the Princeton \depm~\cite{BCFKMN14} based on the Iowa Electronic Market~\cite{iem}. Augur switched from automated book making to splitting in v2~\cite{AKPWZ19}, Gnosis implemented splitting in CTF~\cite{ctf}, and \depms built on CTF, including Polymarket~\cite{poly} and Omen~\cite{Omen_docs}, use it.

\begin{definition}[Splitting and merging]\label{def:splitmerge}
Fix a market $M=(\Omega,J,R,t_{\mathrm{res}})$ in numeraire $\mathcal{N}$. A \emph{splitting/merging} mechanism provides counterparty-free state transitions for a fixed \emph{unit bundle} $B$ of WTA token types such that
\[
\sum_{j\in B} R(\omega)_j = 1 \quad\text{for all }\omega\in\Omega.
\]
\emph{Splitting} maps $1\cdot\mathcal{N}\mapsto \sum_{j\in B} j$ (minting one unit of each $j\in B$), and \emph{merging} maps $\sum_{j\in B} j \mapsto 1\cdot\mathcal{N}$ (burning them), typically before resolution. Repeating the operation yields multiple bundles.
\end{definition}

Recall that in a WTA market, exactly one share in a set of shares will payout \$1. This means that holding a complete portfolio of every share is equivalent (in payoff, ignoring fees and discounting) to holding \$1. A splitting gadget allows a trader to input \$1 and receive a portfolio contain 1 share of each outcome. Generally, a \textit{merging} gadget is also available where a complete set can be redeemed for \$1 at any time before the market closes. Alice can obtain a set of shares and list asking prices (through a limit order book or by being the first liquidity provider in an AMM) for some or all of the shares, and if Bob is willing to buy a share from Alice, the first trade occurs. The pros to splitting is that the operator has zero exposure to the market, while the con is that Alice must wait for a second trader, Bob, before she can trade. 

In YNB markets, each outcome’s YES/NO pair is its own two-outcome WTA market. Splitting is per outcome: converting \$1 of collateral mints one $j_{\mathsf{k_Y}}$ and one $j_{\mathsf{k_N}}$ for the chosen $k$. Across outcomes, supplies are uncoupled---the total minted for the outcome $\mathsf{A}$'s bundle need not match that for outcomes $\mathsf{B}$ or $\mathsf{C}$. 

\RQ{If the first trader makes one dollar, someone must lose a dollar. It either needs to be the operator or a second trader. This forces \depms to choose between needing a loss fund or having no instant liquidity for the first trader.}

In the market initialization component of the modular workflow, a third approach, \textit{matching}, was used by InTrade~\cite{intrade} and a variant by Fairlay~\cite{fairlay}, and is used today by Kalshi. 

\begin{definition}[Matching]\label{def:matching}
Fix a market $M=(\Omega,J,R)$ in numeraire $\mathcal{N}$. For each label $j\in J$, define its per-unit worst-case payoff
\[
c_j \;:=\; \sup_{\omega\in\Omega} R(\omega)_j,
\]
and define the complementary label $\bar{j}$ with payoff
\[
R(\omega)_{\bar{j}} \;:=\; c_j - R(\omega)_j.
\]
A \emph{matching} mechanism accepts orders of the form $(j,q,p)$, meaning ``trade $q\ge 0$ units of label $j$ at price $p$ (in $\mathcal{N}$) per unit,'' and executes a trade only when a buy order and a sell order coincide on $(j,q,p)$. Upon such a match, the mechanism:
(i) transfers $qp$ units of $\mathcal{N}$ from the buyer and $q(c_j-p)$ units of $\mathcal{N}$ from the seller into escrow;
(ii) mints $q$ units of label $j$ to the buyer and $q$ units of label $\bar{j}$ to the seller.
At resolution $\omega\in\Omega$, the two positions pay $qR(\omega)_j$ and $qR(\omega)_{\bar{j}}$, which sum to $qc_j$, exactly covered by the escrow; hence the operator is riskless.
\end{definition}

Briefly, it mirrors a futures market, where Alice posts a desired short/long position at a chosen price on an orderbook with a margin account holding enough cash to cover her maximum loss if she obtains the position. If Bob is willing to take the other side, also with sufficient margin for his maximum loss, the operator matches them, creates two shares and gives them to Alice and Bob. Alice and Bob are not counterparties, both settle with the operator once the shares are created, however their coincidence of wants (COW) is necessary for the operator to create shares at no risk to itself. 

\RQ{A \depm operated through smart contracts should use automated bookmaking or splitting/merging. Matching effectively requires a central limit order book, which is too expensive to run on-chain at scale~\cite{MoCl23}.}

\subsection{Trading}\label{wf:trade}

Once outcome shares are in circulation, they can be traded any way fungible blockchain tokens can be traded. This includes \textit{centralized exchanges} that custody the tokens and use central order book (CLOBs); \textit{partially decentralized exchanges} where CLOB matching is done off-chain and settlement is done on-chain; or \textit{fully on-chain exchanges}, of which automated market makers (AMMs) are the most common. Recent research has identified a failure mode in hybrid order books where orders matched off-chain can subsequently fail during on-chain settlement. Attackers can exploit this gap by selectively invalidating orders, illustrating an important security trade-off in hybrid execution~\cite{SJWWC26,CLS26}.

Of interest, AMMs were born out of Gnosis' research into automated bookmarking for prediction markets. They first developed alternatives to the LMSR rule, including the constant product rule~\cite{LK17}. In parallel, Bancor worked on exchanges where multiple traders could contribute tokens to a common liquidity pool. Uniswap v1 merged these two ideas to create the basic template of an AMM that is common today.

Circling back to automated bookmaking, splitting can be used in conjunction with AMMs to realize automated booking, as first used in Gnosis' closed beta Sight~\cite{Bar19}. An operator begins with $2n$ units of the numeraire, \eg $2n$ USDC. It takes half of the money, $n$ USDC, and uses the splitting gadget to obtain $n$ shares in each outcome. For each outcome share, it estimates the outcome probability as $p_i$ and starts an AMM with initial liquidity of $n$ outcome shares and $p_i\cdot n$ USDC. It repeats for each outcome. Thus assuming the probabilities sum to 1, the operator has split its remaining $n$ USDC across the AMMs. The operator can lose up to $2n$ USDC in divergence loss, assuming traders can split as well (and thus create the shares necessary to drain the AMM of its USDC), in addition to failing (through stale prices) to realize profits (\ie LVR~\cite{MMRZ22}). The important point is that the operator is still solvent per Definition~\ref{def:axioms} as the $2n$ USDC is either locked in the splitting gadget or locked in the AMM.

Despite the direct lineage between prediction markets and AMMs, AMMs are problematic for prediction market trading. \depm outcome shares behave in specific ways that differ from typical crypto-assets and tokens. The price is strictly bounded between \$0 and \$1, the value of a share can jump to \$0 or \$1 near-instantly when an event outcome is finalized, and once finalized, the price is permanent. When real world events occur, AMMs can be drained faster than liquidity providers can withdraw liquidity. 

\gap{Adapting AMMs to the unique constraints of predictions markets is at an early stage. Paradigm’s pm-AMM tapers liquidity as a scheduled expiry approaches, which helps for events that crystallize over a fixed time horizon~\cite{MR24}; but many markets jump or resolve unexpectedly, falling outside of the model used for the results.}

When trading on-chain, miners and other users can front-run transactions, an area of study called maximum extracted value (MEV)~\cite{BBDGJKLZ19}. Although the term MEV did not exist at the time, the Princeton \depm describes the MEV problem extensively~\cite{BCFKMN14} and proposes a mitigation now called a frequent batch auction (FBA) (again, the term FBA was not popularized until later~\cite{BCS15}).

\gap{Can we measurement  MEV extraction for on-chain \depms? What mechanisms for order-flow privacy apply to on-chain \depms?}

A final trading-relevant subject for prediction markets is arbitrage. Arbitrageurs ensure market prices are consistent, for example across all shares in a WTA market or between Yes/No bundles in a YNB market. A recent paper studies combinatorial arbitrage on Polymarket between markets with logically related predicates (\eg Republicans win the presidency; Trump wins the presidency) and measures roughly \$40M USD of realized arbitrage profits over the measurement period~\cite{SGKS25}. 



\subsection{Market Resolution}\label{wf:close}

Market resolution approaches specify how a DePM determines and finalizes the winning outcome and triggers settlement. Designs vary in where the ground truth comes from (market prices, external sources, or participants), who can influence the decision, and what incentives/penalties secure it. In this section we review: 
 oracle-less designs (on-chain truth, self-settling markets, auto-resolve rules), using a designated arbiter, using a network of reporters, and crowdsourcing a vote. The term \textit{oracle} is commonly used for any of the latter three approaches.  

\subsubsection{Oracle-less.}

A market topic might reference \textit{on-chain truth} (\eg total value locked in a DeFi service at the end of the year) in which case resolution can be made directly. A \textit{self-settling market}~\cite{BMR17} works with \textit{splitting} and \textit{merging} (see \S\ref{wf:trade}) of shares; it relies on participants with winning shares to purchase the losing shares (for close to \$0) and redeem \$1 by merging them. If losing shares do not trade near \$0 or become illiquid, the broader market might accept winning shares as a substitute for dollars. If a market outcome is contentious with no recognized winner (\eg a poorly defined market in \S\ref{wf:topic}), the market will not settle. An \textit{auto-resolve rule} could also be used such that outcomes are finalized when, say, their token trades above \$0.99 for at least $t$ time. Auto-resolve rules are subject to market manipulation (losing shares are wash traded near \$1 long enough to finalize, in which case they are worth \$1) or griefing attacks (yes shares are traded below \$0.99 to prevent finalization). While an auto-resolve rule is too fragile to rely on solely, it could be used optimistically to provide an initial settlement; only disputed cases would be escalated to an oracle~\cite{Aug15}. 

\subsubsection{Designated arbiters.}

A \cepm will decide events within the platform~\cite{kalshi}. A \depm might outsource decisions to a trusted organization, such as the Associate Press (AP) which experimented with writing US election results into Ethereum.\footnote{Etherscan: \href{https://etherscan.io/address/0x0792724900B551d200D954a5Ed709d9514d73A9F}{AP U.S. Presidential Race Oracle}} From a trust perspective, this is equivalent to using an AP \cepm, but from a logistics perspective, it is simpler for AP to be an oracle than to run an entire \cepm. Further, it allows \textit{trust agility} where AP can be replaced by a competing source with minimal disruption~\cite{BCFKMN14}. A \depm might also used a designated party only as a last resort if other methods lead to a contentious result~\cite{Rey24}. 

Recently interest is in evaluating arbitration through an AI system~\cite{chaos}. A 2025 Chainlink study poses 1,660 Polymarket questions to LLMs and reports ~89.3\% accuracy (sports 99.7\%, politics 84.3\%)  with temporal-reasoning challenges~\cite{ZVW25}. AI can also be used in agent-mode, tasked with actively gathering evidence and reporting if resolution can be reached~\cite{chaos}. AI oracles are potentially vulnerable to adversarial manipulation and poisoning attacks~\cite{ZGWJ25}, and may recall incorrect facts that are incorrectly stated in the training data. An over-reliance on AI systems in generating news, trading in markets based on news, and settling markets based on news could result in reflexivity and feedback loops. 

\subsubsection{Reporter networks.}

Some oracle systems focus on quickly reporting on updating values~\cite{BCCJM21}, such as an asset price or a sports score which could naturally settle some types of prediction markets, such as rapid markets from Augur Turbo~\cite{aug25} and Polymarket~\cite{polylink}. Generally a set (or network) of arbiters (or reporters) will report the value and an aggregation method (\eg median value or a mean value after outlier filtering for quantitative data, mode for qualitative) will update the value on-chain~\cite{ClEsGS21}. Updates are emitted when deviation or heartbeat thresholds are met, giving explicit control over latency/recency and bounding staleness under network stress. A \depm can read the aggregator at a fixed resolution time and settle without discretionary intervention~\cite{aug25,thalesmarket_docs}. Even when an oracle faithfully reports an external data source, the source itself may be manipulated. Recent research finds evidence consistent with settlement-window price manipulation in short-duration Bitcoin prediction markets, illustrating an attack that does not require compromising the oracle~\cite{DJY26}.

\subsubsection{Crowdsourced vote.}

A final mechanism is to determine off-chain facts by polling a set of willing participants~\cite{But13,Sz14,AKPWZ15,AKPWZ19,uma,kleros}. This procedure is defined by four design choices:

\begin{enumerate}
\item \textit{Electorate}: participants stake governance tokens to enrol and votes are either held with all participants, or with a randomly selected subset of all participants.
\item \textit{Ballots:} may be visible to other participants or may be hidden until voting closes (typically with a two-round commit-reveal protocol but more sophisticated methods could be used~\cite{GSZB23}).
\item \textit{Aggregation:} votes are weighed by the amount of staked tokens, categorical questions are decided by plurality (perhaps with minimum quorum) while quantities with a median or smoothed mean. 
\item \textit{Incentives:} participants who affirmed the final result in their individual votes receive rewards, while those who did not will at least forgo rewards, but might also have their participation tokens slashed. 
\end{enumerate}

The precise parameters of these systems vary significantly across many different designs, while deployed systems like UMA's data verification mechanism (DMV)~\cite{uma} and Kleros~\cite{kleros} have settled billions of dollars. Topics like fee mechanisms and proof of stake consensus protocols have received much academic attention, but incentive-compatible designs and formal proofs for oracles is largely under-explored. By contrast, a growing theoretical line proposes truth-elicitation schemes with formal guarantees~\cite{Pre04,SKC25}. However in practice, uptake is absent, in part because these designs are complex to run, sensitive to Sybil attacks, and can fail if voters rely on a common source of information.

Token votes for \depms are hostile environments where an attacker’s external payoff from a wrong resolution exceeds the system’s economic security, which can lead to failure~\cite{FoBo19}. Rational voters are incentivized to agree with the majority, rather than necessarily reporting the ground truth (a situation colloquially known among economists as a Keynesian beauty contest). The system often works because the ground truth is an obvious point where participants will coordinate (known as a Schelling point), however voters might fear voting for truth if large token holders (whales) visibly stake on a wrong outcome early in the vote. An example\footnote{\polyurl{Ukraine agrees to Trump mineral deal before April?}{https://polymarket.com/event/ukraine-agrees-to-give-trump-rare-earth-metals-before-april}} occurred on Polymarket~\cite{Rey25}, leading to concerns of token concentration within the UMA token. At the time of writing, the top five UMA wallets UMA control ~45.6\% of votes and the top 13 exceed 65\%, enough to meet the passing threshold.\footnote{Dune Analytics: \href{https://dune.com/uma\_protocol/uma-protocol}{\$UMA Voting Participation}} If successful attacks depress token prices, the attacks become cheaper, potentially creating a spiral (\cf~\cite{KM21}). 

These concerns affirm design decisions like a secret ballot, using subsets of participants in a vote (potentially split by subject expertise), and redrawing larger panels on appeal with higher bonds/fees to raise capture costs only when disputes persist.

\gap{Deciding market outcomes is a uniquely \depm problem (since a \cepm can arbitrate its own markets). Can we bring a theoretical lens to deployed oracle voting systems and understand what formal guarantees about incentives and safety can be made?}

A \depm will probably use a \textit{hybrid} or \textit{chained approach} that does not rely on a single method. For example, an arbiter from an allowlist might be able to propose a market outcome while also allowing disputes from anyone. If an outcome is disputed, it is escalated to a crowdsourced vote. If the vote is considered defective, a further escalation could allow an Admin account to overrule the decision. The ultimate backstop is the law which would not stop a wrong outcome but could be used to remunerate parties damaged by it.


\subsection{Settlement}\label{wf:clear}

Once the market is resolved, a \depm will enable each winning share to be converted into 1 unit of the numeraire (\eg 1 USD in a stablecoin) and transferred into the user's self-custody. While a \depm in theory could \textit{push} payouts to users, it is common to wait for the user to initiate the redemption---a \textit{pull} mechanism~\cite{ctf}. In this case, users pay the gas cost of the redemption which requires users to hold the native currency of the underlying blockchain. Polymarket offers gasless withdrawals (using OpenGSN relayers~\cite{gsn}), however users can bypass this at their choice. In a pull model, some users may not redeem their shares in a timely manner---a \depm may opt to sweep this surplus into its own capital or burn it, but \depms generally hold it on-chain in perpetuity. As with any smart contract allowing withdraws based account balances, hardening against reentrancy attacks is critical.

The slogan `sweat the game, not the payout~\cite{Fu24}' is used to differentiate regulated sportsbook operators from `neighbourhood bookies,' however even legitimate operators in the US have allegedly denied payouts using a legal loophole.\footnote{A law protects sports books from obvious errors, like a `fat finger' when setting odds, but can be abused to claim long-shot bets were mistakes if they pay out~\cite{Fu24}.} The advantage of a \depm in this context is two-fold: (i) payouts are fully (or largely) autonomous, not subject to human discretion, and (ii) the share structure ensures the operator has zero risk (or predetermined bounded risk, in the case of automated bookmaking) and is therefore financially indifferent to making any fair payout (see the solvency axiom in Definition~\ref{def:axioms}). 


\RQ{A prediction market has no need to fight against paying out a bet the way bookmakers have in practice~\cite{Fu24}. When operations are on-chain, solvency can be demonstrated directly from blockchain data.}


\subsection{Archiving}\label{wf:archive}

Publicly accessible \depm data provides society with a useful forecasting tool, and archival datasets enable calibration, insight into historical events, and replication of findings. On-chain records (state, logs, and calldata) inherit strong archival and verification properties so long as the chain persists. These records can be replayed and exposed via deterministic chain indexers (\eg The Graph, SubQuery) or managed subgraph hosting (\eg Goldsky). Human-readable materials can be stored in content-addressed, peer-to-peer systems (\eg IPFS) and mirrored to permanence layers with economic durability (\eg Arweave, Filecoin), with their content identifiers anchored on-chain. Privately held operational data can be released as signed public snapshots or made queryable via open APIs.

Two kinds of artifacts matter most. First, the market semantics: the market topic, resolution rules, and any clarifications. Platforms such as Augur and Polymarket record stable on-chain identifiers (e.g., market/condition and token IDs) and keep the human-readable documents in content-addressed storage, with their content hashes/CIDs referenced on-chain. Second is the market data, including trading data (time, volume, price), outcome share supply and redemption totals, and timestamps for the status of the market (\eg opened, resolved, finalized). In practice, settlements and token movements are emitted on-chain, then replayed by deterministic chain indexers into queryable tables for research and UX. The same applies to the resolution process, its dispute trail and finalized outcome. If trading is off-chain (fully or partially), \depms will need to expose application indexes/APIs for fast access and publish signed public snapshots for reproducibility. This missing data includes detailed trading, order-book depth, and liquidity metrics.

The following example describes Polymarket's original CTF-based architecture, before its 2026 protocol redesign~\cite{+polyclob,+polyv2}. Polymarket settles markets on the canonical ledger using the UMA Optimistic Oracle (the dispute game) and writes the final payout vector into the Gnosis Conditional Tokens Framework (CTF) (the settlement contract); outcome shares are ERC-1155 tokens (a multi-token standard), so transfers, mints, and redemptions are visible in logs. Trades are matched off-chain on a central limit order book (CLOB) but settle on-chain, while detailed order-book depth, quotes, and liquidity metrics are exposed via Polymarket’s API (called Gamma). For reproducible research, the same on-chain events are also mirrored by deterministic chain indexers (subgraphs run by services such as Goldsky/The Graph), and API time series can be cross-checked against transaction hashes on the ledger.

\gap{A more comprehensive tool could use on-chain data, API data, and external data (news, social media, \etc) to replay a past market, allowing researchers to step forward and backward at each timestamp. It was onerous for us to manually study how markets incorporate news (real and fake) into prices in \S\ref{app:hbo} without such a tool. It would also make it possible to backtest and train trading strategies using strictly point-in-time information, avoiding hindsight bias from revised data or retrospectively curated search results. Update: Recent work has begun addressing this gap through integrated datasets and historical replay environments~\cite{JZZ+26,YDZ+26,YWA+26}. These contributions provide useful components, although a comprehensive system integrating trading, resolution, and contemporaneous external information remains an open opportunity.}

\gap{A useful feature that does not exist yet would add an automated feedback loop after markets resolve. The system could analyze what happened (using on-chain data, platform logs, and external signals) and use it to tune future markets, such as updating fees, incentives, liquidity settings, or oracle parameters. Today these updates are mostly manual or governance-driven.}


\section{Further Discussion}

\paragraph{Composability.} Perhaps the biggest evolution in \depm design is infrastructure~\cite{Gun24}. Early \depms were monolithic, single-vendor codebases. Modern \depms are built from existing infrastructure, which composes through highly standardized interfaces. For example, Polymarket's original architecture was built on Gnosis' Conditional Tokens Framework (CTF)~\cite{ctf}. The numeraire is Circle's USDC stablecoin, which can be bought with a credit card through MoonPay. Trading outcome shares and USDC works out-of-the-box on any platform (on- or off-chain) that supports ERC-1155 tokens. Market outcome disputes are escalated to UMA's DVM oracle~\cite{uma}. Polymarket also uses third party services for bridging assets, embedded wallets (based on email verification), and EIP-3009 gasless withdrawals. Beyond software engineering benefits, building a service by composing modules can enhance trust agility, which is the ability to quickly swap out modules that are faulty or malicious. For example, Polymarket could switch from UMA to say Chainlink or Kleros, with less effort than if the oracle service was vertically integrated. Composability also introduces additional security dependencies and expands the attack surface. For example, in 2026, a compromised third-party component of Polymarket's frontend was used to execute unauthorized transactions, illustrating how a failure in one component can compromise the security of the overall system without exploiting its underlying smart contracts.

\paragraph{Regulation.} In jurisdictions like the United States, both \cepms and \depms are tightly controlled and not always legal to operate~\cite{Du19,Ma24}. The paramount concerns of regulators are not specific to prediction markets or the difference in their microstructure from other forms of wagering (although market integrity is always a concern); rather, regulators control any kind of wagering on political events (a stance challenged by leading economists~\cite{AFGH+08}), as well as any kind of financial service offered to unaccredited consumers. For advocates of liberal markets, \depms are seen as a way to side-step regulation, particularly when the \depm is decentralized in most or all dimensions of our taxonomy.

\section{Conclusion}

Researchers and builders have used a set of shifting designs, definitions, and vocabulary for \depms. We aim to provide a modular framework that is useful for careful comparison between systems with different design choices, showcasing the full set of choices available, uncovering unsolved research problems, understanding the history of \depm ideas, and providing a learning resource for those wanting to catch up on \depms. Our taxonomy does not identify a single best design but helps illustrate the trade-offs between them.


\subsubsection*{Acknowledgements.} We thank Richard Robertson and the reviewers who helped to improve our paper. J. Clark acknowledges support for this research project from (i) the National Sciences and Engineering Research Council (NSERC), Raymond Chabot Grant Thornton, and Catallaxy Industrial Research Chair in Blockchain Technologies, (ii) the AMF (Autorité des Marchés Financiers), and (iii) NSERC through a Discovery Grant. 


\bibliography{bib/depm.bib}


\clearpage
\appendix


\section{List of \depms}
\label{app:list}


\begin{enumerate}

\item \textbf{Bets of Bitcoin} (2011--2014): Centralized event wagering system with BTC as numeraire. Promoted as a prediction market but mechanically it used parimutuel betting. Went offline unannounced with some user funds stuck.

\item \textbf{BitBet} (2012--2020): Centralized event wagering system with BTC as numeraire. Promoted as a prediction market but mechanically it used parimutuel betting. Disruption in 2016 and winddown in 2020. 

\item \textbf{Predictious} (2013--?): Centralized prediction market and CLOB with BTC as numeraire. Promoted as InTrade successor. Appears abandoned in late 2010s.

\item \textbf{Fairlay} (2014--present): Centralized prediction market with BTC as numeraire. Traders were matched on a CLOB (back/lay mechanism). After ownership changes, still operating as Bitcoin Betting. 

\item \textbf{BetMoose} (2014--present) Centralized event wagering system with BTC as numeraire. Promoted as a prediction market but mechanically it used parimutuel betting or back/lay. Still active. 

\item \textbf{Truthcoin / Bitcoin Hivemind} (2014): Decentralized prediction market (\depm) design with follow-up refinements with some code artifacts. Inspired other \depms (particularly around automated book making and token vote market resolution) and sidechain technologies.

\item \textbf{`Princeton' \depm} (2014): Decentralized prediction market (\depm) design as an academic paper only. Inspired other \depms (particularly around share splitting/merging), on-chain CLOBs (frequent batch auctions) and the concept of MEV.  

\item \textbf{Augur} (2015--present): \depm whitepaper design, later deployed on Ethereum (live in 2018) and Polygon (2021). Numeraire is DAI and later USDC. Native token (REP) used in market resolution was one of Ethereum's first ICOs (2015). Still active. Front-ends for Augur include \textbf{Predictions.Global}, \textbf{Veil}, \textbf{Gueser}, and \textbf{Helena Network}.

\item \textbf{BitShares Prediction Markets} (2015--present): \depm functionality was added to the BitShares 2.0 blockchain to support WTA markets and hybrid on/off-chain CLOBs. \depm functionality dormant. BitShares itself is still active but usage has heavily declined.
 
\item \textbf{Gnosis} (2015--present): \depm whitepaper design for Ethereum. Ran closed beta (sight.pm). Pivoted to developing underlying infrastructure for \depms, including the widely used Conditional Tokens Framework (CTF). Co-built Omen based on CTF. Also known for self-custody wallets (Gnosis Safe) and AMM-relevant research (proposing the constant product market maker). Still active. 

\item \textbf{Stox} (2017--2018): Centralized prediction market and CLOB with custom ERC20 token STX as numeraire. Known for celebrity promotions. Abandoned around 2018 after legal issues. 

\item \textbf{Delphy} (2017--?): Prediction aggregator deployed on Ethereum for mobile devices based on points/leaderboard (`play money') rather than money. Appears to have been abandoned within 2-3 years.

\item \textbf{Bodhi} (2017--?): \depm deployed on Qtum blockchain and later Ethereum. Appears to have been abandoned within 2-3 years. 

\item \textbf{BlitzPredict} (2018--2019): Prediction aggregator on Ethereum that appears to have been abandoned before being developed into a full \depm.

\item \textbf{SportX} (2018--): Sports-centric \depm deployed on Ethereum, Polygon, and later its own EVM chain (SX Network). Still active.

\item \textbf{BetProtocol} (2018): Toolkit for \depm infrastructure using custom ERC20 token BEPRO. Bepro Network pivoted in 2021 and no active development on toolkit since.

\item \textbf{Sharpe Capital} (2017--2019): Prediction aggregator on Ethereum that appears to have been abandoned before being developed into a full \depm.

\item \textbf{Amoveo} (\circa 2018--?): \depm functionality into state-channels on a custom PoW L1 chain. Sporadic ongoing development.

\item \textbf{SEER} (2018--?): \depm deployed on custom Graphene-based DPoS L1 chain with custom SEER token as numeraire. Appear abandoned as of 2020.

\item \textbf{Veil} (2019): \depm front-end built on Augur and 0x. Launched and shut down within 2019.

\item \textbf{PredIQt} (2019--?): \depm deployed on EOSIO with IQ/EOS as numeraire. Project pivoted to encyclopedia IQ.wiki and PredIQt appears inactive.

\item \textbf{Catnip.exchange} (2019): \depm front-end for Augur v1 that composed with Balancer AMMs for trading outcome shares. Discontinued after 2020 US presidential election.

\item \textbf{Flux Protocol} (2019--2022): \depm deployed on Ethereum and later NEAR. Appears dormant after 2022. 

\item \textbf{Thales} (2019--2022): Event wagering system deployed on Ethereum and later NEAR. Market resolution with Chainlink. Largely dormant after 2022.

\item \textbf{Omen} (2020--present): \depm built with Gnosis CTF, deployed on Ethereum and later Gnosis Chain (xDai) and Polygon with any ERC20 as numeraire. Market resolution with Reality.eth and Kleros. Still active.

\item \textbf{Polymarket} (2020--present): \depm built with Gnosis CTF, deployed on Polygon with USDC as numeraire. First \depm to receive wide mainstream coverage in the media. Market resolution with UMA. Still active but restricted in some jurisdictions (including the US). 

\item \textbf{PlotX v1} (2020--2022): \depm deployed on Ethereum and later Polygon. Specialized for crypto price predictions. PlotX still active but 2022 pivot left \depm functionality dormant. 

\item \textbf{Reality Cards} (2020--2022): \depm deployed on Ethereum and later Gnosis and Polygon. Outcomes shares are NFTs that can be rented with payouts based on how long a user held the winning NFT (time-weighted to compensate early traders more). Appears abandoned in 2022.

\item \textbf{Prosper} (2021--present): \depm deployed on Avalanche and BSC. Still active.

\item \textbf{Zeitgeist} (2021--present): \depm deployed into the logic of a parachain (custom L1) in the Polkadot/Kusama ecosystem (L0). Still active.

\item \textbf{Polkamarkets} (2021--present): \depm toolkit for EVM chains like Polygon and Moonriver with any ERC-20 as numeraire (and custom ERC20 POLK for governance). Still maintained.

\item \textbf{Hedgehog Markets} (2021--?): Event wagering platform deployed on Solana with USDC as numeraire. Supports both no-loss contests and prediction markets. Appears dormant after 2022.

\item \textbf{Unihedge} (2021): \depm design for EVM with an experimental prototype. Outcomes shares are structured different than a typical prediction market (lots implementing what is known as a Harberger-tax).

\item \textbf{Mojito Markets} (2022--present): \depm designed for Aptos but not yet deployed.

\item \textbf{Insight Prediction} (2024--present): \cepm with blockchain-based payments in various stablecoins, including USDC. Still active.

\item \textbf{Moonopol} (2024--present): \depm deployed on Solana with USDC as numeraire. Still active.

\item \textbf{Miscellaneous}: There are decentralized event wagering systems (or toolkits) that deploy betting structures different from prediction markets. For the earliest systems using such an adjacent approach, we have included them above. However we do not expand on every follow-up project. These include: \textbf{Wagerr}, \textbf{BetDEX}, \textbf{Monaco Protocol}, \textbf{Peerplays}, \textbf{DexWin}, \textbf{DuelDuck}, \textbf{BetterFan}, \textbf{Oriole Insights}, \textbf{BetSwag.gg} and \textbf{Azuro}. We also note here the most prominent \cepms: \textbf{Kalshi}, \textbf{PredictIt}, \textbf{Futuur}, and \textbf{Manifold}. 

\end{enumerate}

\end{document}